\documentclass[a4paper,twocolumn,english,aps,pra,reprint,superscriptaddress,showpacs,showkeys,longbibliography]{revtex4-1}
\usepackage{amsmath,amsfonts,amssymb}
\usepackage{graphicx,subfigure}
\usepackage[colorlinks,urlcolor=blue,linkcolor=blue,anchorcolor=blue,citecolor=blue]{hyperref}
\usepackage{placeins}
\usepackage{color}

\newcommand{\up}{|\!\uparrow\,\rangle}
\newcommand{\down}{|\!\downarrow\,\rangle}

\begin{document}
\title{Scalable nuclear-spin entanglement mediated by a mechanical oscillator}
\author{Puhao Cao}
\affiliation{School of Physics, Huazhong University of Science and Technology, Wuhan 430074, China}
\author{Ralf Betzholz}
\email{ralf\_betzholz@hust.edu.cn}
\author{Jianming Cai}
\affiliation{School of Physics, Huazhong University of Science and Technology, Wuhan 430074, China}
\affiliation{International Joint Laboratory on Quantum Sensing and Quantum Metrology, Huazhong University of Science and Technology, Wuhan 430074, China}		

\begin{abstract}
We propose a solid-state hybrid platform based on an array of implanted nitrogen-vacancy (NV) centers in diamond magnetically coupled to a mechanical oscillator. The mechanical oscillator and the NV electronic spins both act as a quantum bus and allow us to induce an effective long-range interaction between distant nuclear spins, relaxing the requirements on their spatial distance. The coherent nuclear spin-spin interaction, having the form of an Ising model, can be maintained in the presence of mechanical damping and spin dephasing via a pulsed dynamical decoupling of the nuclear spins in addition to the microwave driving field of the electronic spins. The present hybrid platform provides a scalable way to prepare multipartite entanglement among nuclear spins with long coherence times and can be applied to generate graph states that may be used for universal quantum computing. 
\end{abstract}

\date{\today}	
\maketitle

\section{Introduction}
\label{sec:introduction}
One of the most critical challenges in the development of quantum information processing, including quantum computation and quantum simulation, is scalability, which poses stringent requirements on both the interactions between distant qubits and their coherence times. Among the many promising candidates for the implementation of qubits in quantum information processing are solid-state spins, such as the electronic or nuclear spins of nitrogen-vacancy (NV) color centers in diamond. Some of their big advantages are easy preparation and readout of both the electronic and nitrogen nuclear-spin states~\cite{Neumann10a,Childress06,Dutt07}. Also, despite their solid-state environment they can show long coherence times~\cite{Balasubramanian09,Siddharth16}. However, regardless of their superb controllability, NV center spins as building blocks for a scalable quantum processor suffer from a considerable drawback; that is, their magnetic dipole-dipole interaction dramatically decreases with distance, imposing strict conditions on their spatial separation in order to exhibit a non-negligible coupling strength. The concomitant challenge with the weak interaction among distant NV center spins is the necessity of exceedingly long coherence times. Here, one approach to nevertheless ensure the scalability is to explore the possibility of optically coupling NV center arrays in photonic crystal cavities and waveguides~\cite{Riedrich15,Schukraft16,Gould16,Schroder17}.

Another way to overcome these challenges is to consider hybrid platforms including both mechanical and spin degrees of freedom, which offer the potential to couple spins in solid-state systems indirectly via the vibrational mode of a mechanical oscillator, which acts as a long-range mediator~\cite{Rogers14}. In this context, an interaction between the NV center electronic spins and mechanical elements such as cantilevers~\cite{Rabl09,Rabl10,Arcizet11,Kolkowitz11,Yin15}, membranes, graphene sheets~\cite{Reserbat16}, and carbon nanotubes~\cite{Peng16} can be considered. On the other hand, compared with the electron spins of NV centers, the nitrogen nuclear spins show some considerable advantages due to their excellent coherence times. It would therefore be particularly desirable to  realize a long-range interaction between nuclear spins in a scalable NV center array.

Here, we propose a hybrid system, consisting of an array of NV centers magnetically coupled to a mechanical oscillator, in which such an effective long-range nuclear spin-spin interaction is established by two mediators. The mechanical oscillator couples the electronic spins~\cite{Cao17}, which in turn act as a second mediator and induce an Ising interaction between the nuclear spins ~\cite{Bermudez11}. We show that the effective interaction strength can be high enough to maintain a coherent nuclear-nuclear coupling in the presence of environmental noise for realistic parameters by merely applying a small number of spin-echo pulses on the nuclear spins in addition to the continuous microwave driving field of the electronic spins. The present proposal takes advantage of the mechanical-oscillator-mediated long-range interaction and overcomes the limitations on the distance between NV centers in previous studies~\cite{Bermudez11} and thereby paves the way towards a more scalable implementation.  We demonstrate that the present scheme provides a way to generate large-scale graph states and thus a possible platform for the implementation of universal measurement-based quantum computation~\cite{Raussendorf01,Raussendorf03,Briegel09}.

This paper is organized in the following way. In Sec.~\ref{sec:model} we give an introduction to the hybrid platform we consider. We then show in Sec.~\ref{sec:effective} how this system can generate an effective coupling of the nuclear spins in a suitable parameter regime. Section~\ref{sec:decoherence} is dedicated to the analysis of how this coupling is affected by the decoherence of the individual constituents of the hybrid system. In Sec.~\ref{sec:application} we investigate some possible applications, namely, the generation of multipartite entangled states of the nuclear spins, before we conclude in Sec.~\ref{sec:conclusion}.

\section{Model}
\label{sec:model}
\begin{figure}[t]
	\includegraphics[width=0.9\linewidth]{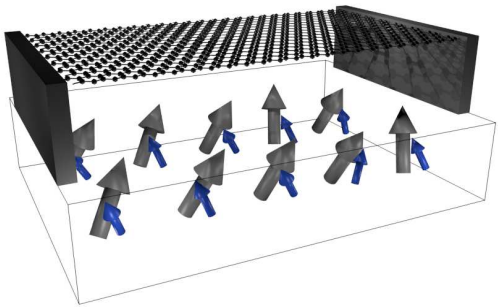}
	\caption{Schematic representation of the hybrid platform we consider (not to scale). An array of nitrogen-vacancy centers in a diamond substrate (here transparent), in which every electronic spin (large, gray arrows) is coupled to one nitrogen nuclear spin (small, blue arrows) and interacts with a mechanical oscillator, such as a membrane carrying a direct current suspended over the diamond surface.}
	\label{fig:1}
\end{figure}

We consider a system of $N$ NV color centers in diamond~\cite{Jelezko06} which are located in the proximity of a mechanical oscillator, e.g., NV centers implanted in a diamond substrate and an oscillator hovering above its surface. Every NV center consists of an electronic spin coupled to the nuclear spin of the nitrogen atom. A schematic representation of the platform we investigate is shown in Fig.~\ref{fig:1}.
In this section we introduce the Hamiltonian of the individual NV centers and their interaction with the mechanical oscillator and make some general considerations before we derive an effective nuclear-spin interaction mediated by both the electronic spins and the oscillator in Sec.~\ref{sec:effective}.

\subsection{NV center Hamiltonian}
In the ground-state triplet of the NV center, the electronic spin of the $i$th center is represented by the spin-1 operators $S^{(i)}_\kappa$, with $\kappa=x,y,z$. Applying a static magnetic field of strength $B$ along the NV center $z$ axis and a microwave field with frequency $\omega_e$ then yields an electronic ground-state Hamiltonian~\cite{Doherty13} 
\begin{align}
H_e^{(i)}=D(S_z^{(i)})^2+\gamma_e BS_z^{(i)}+\sqrt{2}\Omega_i\cos(\omega_e t)S_x^{(i)},
\end{align}
with the zero-field splitting $D=(2\pi)\,2.87\ $GHz, the gyromagnetic ratio $\gamma_e=(2\pi)\,2.8\ $MHz/G, and the Rabi frequency $\Omega_i$. Here, the factor of $\sqrt{2}$ in the Rabi frequency has been introduced for later convenience, and we have set $\hbar=1$. Appropriately choosing the frequency of the microwave field allows us to confine the electronic spin dynamics to a two-state subspace. In order to do so, we assume a driving field resonant with the transition between the states $|m_s=0\rangle$ and $|m_s=1\rangle$, i.e., a driving field frequency $\omega_e=D+\gamma_e B$. The electronic degree of freedom of the $i$th center can then be expressed in terms of the usual Pauli operators $\sigma^{(i)}_\kappa$, with $\kappa=x,y,z$, where we denote the $+1$ and $-1$ eigenstates of $\sigma_z^{(i)}$ as $\up_e^{(i)}$ and $\down_e^{(i)}$, respectively. Formally, the confinement to this subspace is described by the substitutions $S_z^{(i)}\to(1_i+\sigma_z^{(i)})/2$ and $S_x^{(i)}\to\sigma_x^{(i)}/\sqrt{2}$, where $1_i$ denotes the unity operator in the two-state subspace of the $i$th electronic spin. This leads to an electronic Hamiltonian in the frame rotating at the driving-field frequency $\omega_e$ given by
\begin{align}
\label{eq:H_e}
H_e^{(i)}=\frac{\Omega_i}{2}\sigma_x^{(i)}.
\end{align}

Furthermore, the Hamiltonian of the $i$th NV center's nuclear spin is given by 
\begin{align}
H_n^{(i)}=\gamma_n BI_z^{(i)},
\end{align}
with the nuclear spin operators $I_\kappa^{(i)}$, for $\kappa=x,y,z$, and the gyromagnetic ratio $\gamma_n=(2\pi)\,0.43\ $kHz/G. Here, and throughout the subsequent text, we assume NV centers in which the nitrogen atoms are of the isotope $^{15}$N, such that their spin operators can likewise be written in terms of Pauli operators $\tau_\kappa^{(i)}$, for $\kappa=x,y,z$, according to $I_\kappa^{(i)}=\tau_\kappa^{(i)}/2$. In this case the $+1$ and $-1$ eigenstates of $\tau_z^{(i)}$ will be represented by $\up_n^{(i)}$ and $\down_n^{(i)}$, respectively. We briefly mention, that diamond with NV centers comprising $^{15}$N can be readily fabricated by ion implantation and identified by optically detected magnetic resonance~\cite{Rabeau06,Ohno12}. Alternatively, effective spin-1/2 nitrogen nuclear spins of the usual $^{14}$N isotope can be realized by selectively driving one spin transition on resonance, similar to the electronic spin case from above. This would, for our purposes, result in only a slight change of the parameters and an additional driving field. 

The dipole-dipole interaction between the electronic and nuclear spins in an NV center is given by
\begin{align}
\label{eq:d-d}
H_{e-n}^{(i)}=A^{\parallel}S_z^{(i)}I_z^{(i)} +A^{\perp}\left[S_x^{(i)}I_x^{(i)}+S_y^{(i)}I_y^{(i)}\right],
\end{align}
with the longitudinal and transverse coupling strengths $A^{\parallel}$ and $A^{\perp}$, respectively. However, due to the large mismatch between the energy splittings of the electronic spin states ($D\pm\gamma_e B$) and the nuclear-spin states ($\gamma_n B$) one can make a secular approximation and consider only the longitudinal part of the dipole-dipole interaction~\eqref{eq:d-d} while neglecting the transversal part, which accounts for direct spin-flip processes~\cite{Bermudez11,Suter17}. For the sake of convenience we will therefore use the abbreviation $A=A^{\parallel}$ for the longitudinal coupling strength, which we assume is uniform for all centers and whose magnitude is given by $A=(2\pi)\,3.05\ $MHz~\cite{Rabeau06,Felton09}. The ground-state Hamiltonian $H^{(i)}_{\rm NV}$ of the $i$th center in the effective spin-1/2 subspace of the electronic spins then takes the form
\begin{align}
\label{eq:H_NV}
H_{\rm NV}^{(i)}=\frac{\Omega_i}{2}\sigma_x^{(i)}+\frac{\omega}{2}\tau_z^{(i)}+\frac{A}{4}\sigma_z^{(i)}\tau_z^{(i)},
\end{align}
with the effective energy splitting of the nuclear-spin states defined by $\omega=\gamma_nB+A/2$. Here, the energy term $A\tau_z^{(i)}/4$ stems from the confinement of the electronic spin dynamics, i.e., the transition from spins-1 to spin-1/2 operators mentioned before. Choosing NV centers formed by $^{14}$N would additionally result in an electronic energy term $A\sigma_z^{(i)}/4$ that is absent in our case.

\subsection{Interaction with the oscillator}
Up to this point we have not included the presence of the mechanical oscillator in our description. We assume a geometry in which the oscillation of the mechanical element is along the $z$ axis and introduce the annihilation operator $a$ and creation operator $a^\dagger$ associated with the vibrational mode. The oscillation frequency will be denoted by $\nu$. For a mechanical oscillator one can envision novel nanomechanical oscillators~\cite{Ekinci05} such as thin membranes, graphene sheets~\cite{Geim07,Miao14,Sharma15}, and carbon nanotubes~\cite{Laird15} carrying a direct current, thereby emanating a magnetic field that couples to the electronic spins of the NV centers. The coupling to the nuclear spins is here neglected due the fact that $\gamma_n$ is several orders of magnitude smaller than $\gamma_e$. In a linear approximation of the magnetic field's $z$ dependence the presence of the oscillator then leads to an interaction term, 
\begin{align}
\label{eq:e-osc}
H^{(i)}_{e-{\rm osc}}=g_i(a+a^\dagger)\sigma_z^{(i)},
\end{align}
where $g_i$ is the coupling strength between the $i$th NV center and the mechanical oscillator, which is proportional to the gradient of the magnetic field produced by the oscillator along the $z$ direction~\cite{Rabl09,Peng16}. The full Hamiltonian of the hybrid platform, reported in the frame rotating at the microwave driving-field frequencies $\omega_e$ and the nuclear-spin transition frequency $\omega$, is then given by
\begin{align}
\label{eq:H_1}
H=H_0+\sum_{i=1}^N\left[\frac{A}{4}\sigma_z^{(i)}\tau_z^{(i)}+g_i(a+a^\dagger)\sigma_z^{(i)}\right],
\end{align}
with the free Hamiltonian defined as
\begin{align}
H_0=\nu a^\dagger a+\sum_{i=1}^N\frac{\Omega_i}{2}\sigma_x^{(i)},
\end{align}
in which we neglected the zero-point energy of the mechanical oscillator.

\subsection{Remarks}
\label{sec:remarks}
Before we continue we conclude this section with a few remarks. In writing the system's Hamiltonian in the form of Eq.~\eqref{eq:H_1} we have neglected a part that was previously suggested~\cite{Bermudez11} as a quantum bus for the interaction of nuclear spins in diamond, namely, the dipole-dipole interaction between the electronic spins of the different NV centers. Since the dipole-dipole coupling strength drastically decreases with distance, that is to say, is inversely proportional to the distance cubed, in this case the distance between the NV centers has to be in the range of a few tens of nanometers to exhibit a sufficiently strong interaction strength~\cite{Neumann10b}. This makes it a challenging task to find or fabricate a suitable array of NV centers which are situated closely enough to each other in order to exhibit a strong direct coupling and thereby implement an array of coupled nuclear spins. Here, we mention that in many engineered arrays, the inter-NV center distance is on the order of a few hundreds of nanometers~\cite{Chen17,Stephen18}, making the direct dipole-dipole coupling negligible. However, by additionally employing the mechanical oscillator as a long-range mediator of the coupling, we propose a system in which the restrictions on the NV center spatial separation are given by the mechanical oscillators mode function. With vibrational modes whose mode functions cover a large area this provides a suitable platform in terms of the scalability of the array. The direct electronic dipole-dipole interaction may be incorporated in our model straightforwardly, but based on the above arguments we refrain from doing so. 

In addition to the possibility of addressing a particular nuclear-spin transition, applying a resonant radio-frequency driving field of strength $\Omega_n^{(i)}$ for the nuclear spins can, in some cases, also be desirable in order to protect the nuclear spins from environmental noise. This would lead to an additional term $H_{nd}=\sum_{i=1}^N\Omega_n^{(i)}\tau_x^{(i)}/2$ in the Hamiltonian $H_0$, in a frame rotating at the driving frequency. However, since the nuclear-spin coherence time is sufficiently long and we consider the nuclear spin-1/2 isotope $^{15}$N, we abstain from including it in our description. As we will see later in Sec.~\ref{sec:application}, this will also be favorable for the generation of graph states since the driving term would no longer commute with the nuclear spin-spin interaction.

\section{Effective nuclear interaction}
\label{sec:effective}
In this section we will show that in an appropriate parameter regime the hybrid system introduced above induces an effective coupling between the originally uncoupled nuclear spins. This coupling is mediated by the electron spins, whose interaction is mediated by the mechanical oscillator. In a step-by-step elimination of the oscillator and the electronic spins we derive an effective Hamiltonian describing the nuclear-nuclear interaction.

\subsection{Elimination of the electron-oscillator interaction}
In the Hamiltonian~\eqref{eq:H_1} the interaction between the electronic spins and the mechanical oscillator can be seen as a state-dependent constant force on the oscillator. This interaction term can be eliminated readily by applying the polaron transformation~\cite{Mahan00,Wilson02,Agarwal13}
\begin{align}
\label{eq:polaron}
\mathcal{P}=\prod_{i=1}^ND\big(\alpha_i\sigma_z^{(i)}\big),
\end{align}
with the relative coupling strength 
\begin{align}
\alpha_i=\frac{g_i}{\nu}
\end{align}
and the displacement operator $D(\alpha)=\exp(\alpha a^\dagger-\alpha^\ast a)$ of the oscillator. For the above choice of $\alpha_i$ the interaction term~\eqref{eq:e-osc} vanishes, and all residual effects of the oscillator dynamics are solely encoded in the transformation of the microwave driving-field term in $H_0$. These effects can be analyzed by expanding the transform in powers of $\alpha_i$. However, in our case we assume $\alpha_i\ll1$ and a mechanical oscillator which is cooled close to its motional ground state~\cite{Schliesser08,Teufel11,Chan11}, so that during the transformation of the electronic driving field~\eqref{eq:H_e} we make the approximation
\begin{align}
\label{eq:approx_1}
\mathcal{P}\Bigg[\sum_{i=1}^N\frac{\Omega_i}{2}\sigma_x^{(i)}\Bigg]\mathcal{P}^\dagger\approx \sum_{i=1}^N\frac{\tilde\Omega_i}{2}\sigma_x^{(i)},
\end{align}  
with $\tilde\Omega_i=\Omega_i\exp(-2\alpha_i^2)$. In our case it is a reasonable assumption to make the above approximation, and further details on this are given in Appendix~\ref{app:A}. 
The transformation of the full Hamiltonian~\eqref{eq:H_1}, which we will denote by $H_\mathcal{P}=\mathcal{P}H\mathcal{P}^\dagger$, thereby has the form
\begin{align}
\label{eq:H_P}
H_\mathcal{P}=H_0+\frac{A}{4}\sum_{i=1}^N\sigma_z^{(i)}\tau_z^{(i)}-\nu\sum_{i,j=1}^N\alpha_i\alpha_j\sigma_z^{(i)}\sigma_z^{(j)},
\end{align}
where in $H_0$ the electronic Rabi frequencies $\Omega_i$ are replaced by $\tilde\Omega_i$. We see that the oscillator induces a coupling of the magnitude $g_ig_j/\nu$ between the otherwise uncoupled electronic spins $i$ and $j$ that depends only on the distribution of their oscillator coupling strength $g$, not on their spatial separation. We briefly mention that the same Hamiltonian can also be derived by employing effective Hamiltonian theory~\cite{James07,Gamel10} instead of the polaron transformation~\eqref{eq:polaron}.

\subsection{Elimination of the electron-nuclear interaction}
Our aim is still to establish an effective interaction between the nuclear spins. In order to achieve this, following the polaron transformation $\mathcal{P}$ we additionally perform a Schrieffer-Wolff transform~\cite{Schrieffer66,Bravyi11}  
\begin{align}
\label{eq:Schrieffer}
\mathcal{S}=\prod_{i=1}^Ne^{-i\beta_i\sigma_y^{(i)}\tau_z^{(i)}}
\end{align}
of the Hamiltonian~\eqref{eq:H_P}, where we defined the parameter 
\begin{align}
\beta_i=\frac{A}{4\Omega_i}.
\end{align}
As is common for a Schrieffer-Wolff transform, the generator $-i\beta_i\sigma_y^{(i)}\tau_z^{(i)}$ is chosen such that the direct electron-nuclear interaction in Eq.~\eqref{eq:H_P} is eliminated in the first order of $\beta_i$. The magnitude of the longitudinal electron-nuclear coupling and typical values of the electronic Rabi frequency, viz., $A= (2\pi)\,3.05\ {\rm MHz}$ and $\Omega_i\approx (2\pi)\,10-20\ {\rm MHz}$,  imply $\beta_i\ll1$ and suggest a truncation of the transform after the second order of $\beta_i$ by neglecting all higher-order terms. The transformation $\mathcal{S}H_\mathcal{P}\mathcal{S}^\dagger$ of the Hamiltonian~\eqref{eq:H_P} up to second order in $\beta_i$ will be denoted by $H_\mathcal{S}$ and reads
\begin{align}
\label{eq:H_S}
H_\mathcal{S}=H_0-\sum_{i<j}^NJ_e^{(ij)}\sigma_z^{(i)}\sigma_z^{(j)}-\sum_{i<j}^NJ_n^{(ij)}\sigma_x^{(i)}\sigma_x^{(j)}\tau_z^{(i)}\tau_z^{(j)},
\end{align}
where the electronic and nuclear spin-spin coupling constants that we introduced here are respectively given by
\begin{gather}
J_e^{(ij)}=2\alpha_i\alpha_j\left(1-2\beta_i^2-2\beta_j^2\right)\nu,\\
J_n^{(ij)} = 8\alpha_i\alpha_j\beta_i\beta_j\nu.
\end{gather}
In writing the Hamiltonian in the form of Eq.~\eqref{eq:H_S} the electronic Rabi frequencies in $H_0$ have been renormalized again according to $\bar\Omega_i=(1+2\beta_i^2)\tilde\Omega_i$. Although still involving some electronic operators, the last term of Eq.~\eqref{eq:H_S} now constitutes a direct coupling of the individual nuclear spins, as was our aim to establish. This coupling is mediated by both the mechanical oscillator and the electron-nuclear dipole-dipole interaction, and therefore is in second order of both $\alpha$ and $\beta$. Including a nuclear driving field, as mentioned in Sec.~\ref{sec:remarks}, leads to additional terms in $H_0$, as shown in Appendix~\ref{app:B} for the sake of completeness.

\subsection{Effective nuclear spin-spin interaction}
We now have a closer look at the nuclear-nuclear interaction in the Hamiltonian~\eqref{eq:H_S}. Therefore, we first define the eigenstates of the operators $\sigma_x^{(i)}$ and $\tau_x^{(i)}$ according to $|\pm\rangle^{(i)}_\lambda=[\up^{(i)}_\lambda\pm\down^{(i)}_\lambda]/\sqrt{2}$, for $\lambda=e,n$. We now assume that the electronic spins are all equally prepared in either one of the two eigenstates $|\pm\rangle^{(i)}_e$, which can be easily achieved by optical ground-state polarization followed by a $\pi/2$ pulse. By discarding the fast-rotating terms in the electron-electron interaction $J_e^{(ij)}\sigma_z^{(i)}\sigma_z^{(j)}$ in a rotating-wave approximation we find that in this case the electronic and nuclear dynamics are decoupled from each other since the action of every $\sigma_x^{(i)}\sigma_x^{(j)}$ in $H_{\mathcal{S}}$ is the unity operator, and the effective nuclear-spin Hamiltonian is reduced to
\begin{align}
\label{eq:H_eff}
H_{\rm eff}=-\sum_{i<j}^NJ_n^{(ij)}\tau_z^{(i)}\tau_z^{(j)}.
\end{align}
We mention that this effective Hamiltonian has an additional local term $\sum_{i=1}^N\omega\tau_z^{(i)}/2$, which is absent in the rotating frame reported here. At this point we assume a uniform coupling strength $g$ and electronic Rabi frequency $\Omega$ and in the following restrict ourselves to the set of parameters summarized in Table~\ref{tab:parameters} for later reference.
\FloatBarrier
\begin{table}[h!]
	\caption{Choice of parameters}
	\label{tab:parameters}
	\centering
	\def\arraystretch{1.5}
	\begin{tabular}{cccc}
		\hline\hline
		$\Omega/2\pi$ 	& $\nu/2\pi$ & $A/2\pi$    & $g/2\pi$ \\\hline
		$15.25\ $MHz  \quad  & \quad $2\ $MHz	\quad & \quad $3.05\ $MHz \quad & \quad $0.1\ $MHz \\  
		\hline\hline
	\end{tabular}
\end{table}
\FloatBarrier
\noindent
This yields $\alpha=\beta=1/20$, and we thereby find an effective nuclear spin-spin interaction strength
\begin{align}
J_n=8\alpha^2\beta^2\nu=\frac{A^{2}g^2}{2\Omega^2\nu}= (2\pi)\,0.1\ {\rm kHz}.
\end{align}
We see that compared with previous proposals~\cite{Bermudez11}, our scheme relaxes the requirement on the proximity of the electronic spins to each other while keeping the same order of magnitude of the effective coupling strength $J_n$. A nonuniform electron-oscillator coupling strength can, in principle, be compensated by appropriately adjusting the individual Rabi frequencies, in order to ensure an identical nuclear spin-spin interaction strength $J_n$.

\begin{figure}[!t]
	\includegraphics[width=0.95\linewidth]{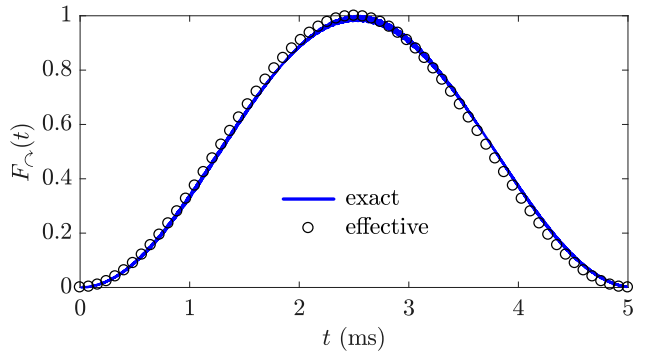}
	\caption{Comparison of spin-flip fidelity $F_\curvearrowright(t)$ under the exact Hamiltonian~\eqref{eq:H_1} (solid line) and effective Hamiltonian~\eqref{eq:H_eff} (circles) for two NV centers. We chose the initial state $|+\rangle_e^{(1)}|+\rangle_e^{(2)}|+\rangle_n^{(1)}|+\rangle_n^{(2)}$ and the mechanical oscillator in its ground state. The parameters are the ones from Table.~\ref{tab:parameters}, resulting in a nuclear spin-flip time of $\pi/2J_n= 2.5\ $ms.}
	\label{fig:2}
\end{figure}

To investigate the effective interaction in more detail, for the moment we considered only two NV centers and have a look at a complete spin-flip processes of the nuclear spins. We assume the electronic degrees of freedom are initially prepared in the subspace where the electronic and nuclear dynamics are decoupled, e.g., the electronic initial state $|+\rangle_e^{(1)}|+\rangle_e^{(2)}$, and consider the mechanical oscillator to be cooled to the ground state. In order to bring out the nuclear spin-spin coupling most clearly, we choose the initial state $|+\rangle_n^{(1)}|+\rangle_n^{(2)}$. The preparation and readout of the nuclear-spin states can be achieved via a quantum nondemolition-measurement scheme~\cite{Neumann10a}. The target state of a nuclear spin flip thereby is $|-\rangle_n^{(1)}|-\rangle_n^{(2)}$, and as a measure of the process fidelity we employ
\begin{align}
F_\curvearrowright(t)=\langle-|_n^{(1)}\langle-|_n^{(2)}{\rm Tr}_{e-{\rm osc}}\{\rho(t)\}|-\rangle_n^{(1)}|-\rangle_n^{(2)},
\end{align}
where $\rho$ denotes the full system's density operator and ${\rm Tr}_{e-{\rm osc}}\{\,\cdot\,\}$ is the partial trace over the electronic and mechanical oscillator degrees of freedom. For a perfect spin flip, as described by the effective nuclear Hamiltonian~\eqref{eq:H_eff}, the time evolution of this fidelity is simply given by $F_\curvearrowright(t)=\sin^2(J_nt)$, resulting in a nuclear spin-flip time $\pi/2J_n= 2.5\ $ms. In Fig.~\ref{fig:2} we show a comparison of the dynamics under the exact Hamiltonian~\eqref{eq:H_1} (numerical propagation) and the effective nuclear spin-spin Hamiltonian~\eqref{eq:H_eff}. 
The fast oscillation of the individual nuclear spins with frequency $\omega$ due to the local terms in the Hamiltonian~\eqref{eq:H_NV}, which are absent in the rotating frame, can be eliminated by applying a spin-echo pulse~\cite{Levitt} after half the evolution time $t$, in order to clearly see the nuclear-interaction dynamics in the original frame~\cite{Bermudez11}. 

The slight discrepancy between the effective dynamics under the effective Hamiltonian and the exact one stems from two aspects. First, we neglected the transformations $\mathcal{P}$ and $\mathcal{S}$ of the initial state; secondly, our Schrieffer-Wolff transform was performed only in second order of $\beta=A/4\Omega$, and the factor $\alpha=g/\nu$ has to be sufficiently small for the elimination of the oscillator. In order to improve the accuracy of the effective Hamiltonian the magnitude of both these factors would have to be decreased, thereby also diminishing the coupling strength $J_n$. This fact shows the trade-off between the choice of the parameters of Table~\ref{tab:parameters} in terms of the achievable nuclear spin-spin coupling. Explicitly, this trade-off entails two aspects. Since the parameter $A$ is an intrinsic property of the NV center, the Rabi frequency $\Omega$ has to be chosen such that $\beta$ is small enough for the approximation yet the effective coupling strength is still acceptable. Second, the effective interaction could naturally be enhanced by a superior coupling strength $g$, which then requires a mechanical oscillator with a higher frequency $\nu$ in order for $\alpha$ to be sufficiently small. However, since even higher electron-oscillator interaction strengths are likely too demanding for experimental implementations we restrict ourselves to values derived in previous proposals~\cite{Rabl09,Peng16}.

\section{Robustness against decoherence}
\label{sec:decoherence}
In a realistic scenario the practical applicability of quantum systems is always limited by the decoherence of the individual degrees of freedom. In this section we will investigate the impact of this decoherence on the effective nuclear spin-spin interaction, as described by the Hamiltonian~\eqref{eq:H_eff} and shown in Fig.~\ref{fig:2}. Despite the fact that the mechanical oscillator, the electronic spins, and the nuclear spins naturally undergo their respective damping and dephasing processes at the same time, we first analyze the influence of their individual decoherence separately in order to identify the main source of infidelity before we give some overall estimate including all three processes at the same time. Since our main goal is the generation of highly entangled nuclear-spin states, for two NV centers and the same initial state $|+\rangle_n^{(1)}|+\rangle_n^{(2)}$ as in Fig.~\ref{fig:2}, we define the target state as the maximally entangled state
\begin{align}
|\Psi\rangle=\frac{1}{\sqrt{2}}\Big[|+\rangle_n^{(1)}|+\rangle_n^{(2)}+i|-\rangle_n^{(1)}|-\rangle_n^{(2)}\Big],
\end{align}
which is created by the effective Hamiltonian~\eqref{eq:H_eff} after the evolution time $\pi/4J_n= 1.25\ $ms. The preparation fidelity
\begin{align}
F_\Psi(t)=\langle\Psi|{\rm Tr}_{e-{\rm osc}}\{\rho(t)\}|\Psi\rangle
\end{align}
of this state is the quantity we will investigate in the following.

\subsection{Mechanical damping}
\label{sec:decoherence_mechanical}
\begin{figure}[!t]
	\centering
	(a)\hspace{-2mm}\vtop{\vskip+1ex\hbox{\includegraphics[width=0.95\linewidth]{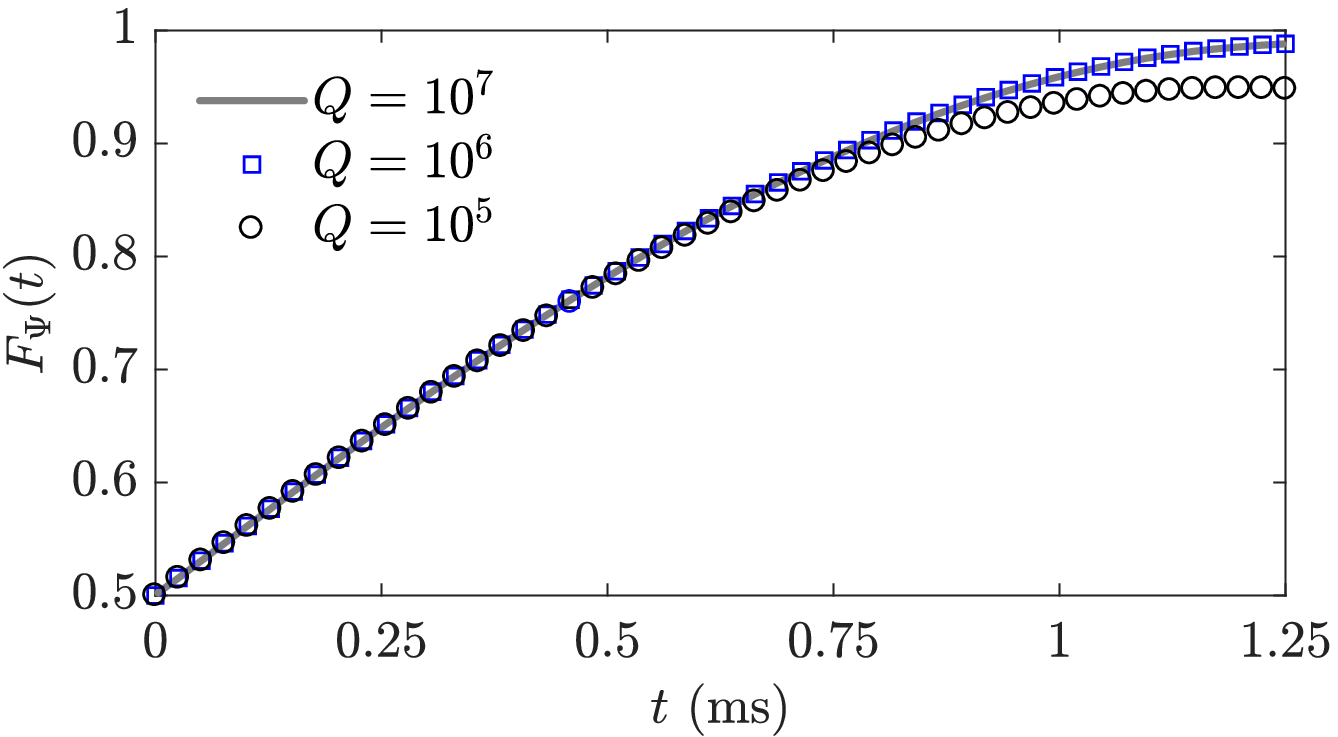}}}\hspace{2mm}
	(b)\hspace{-2mm}\vtop{\vskip+1ex\hbox{\includegraphics[width=0.95\linewidth]{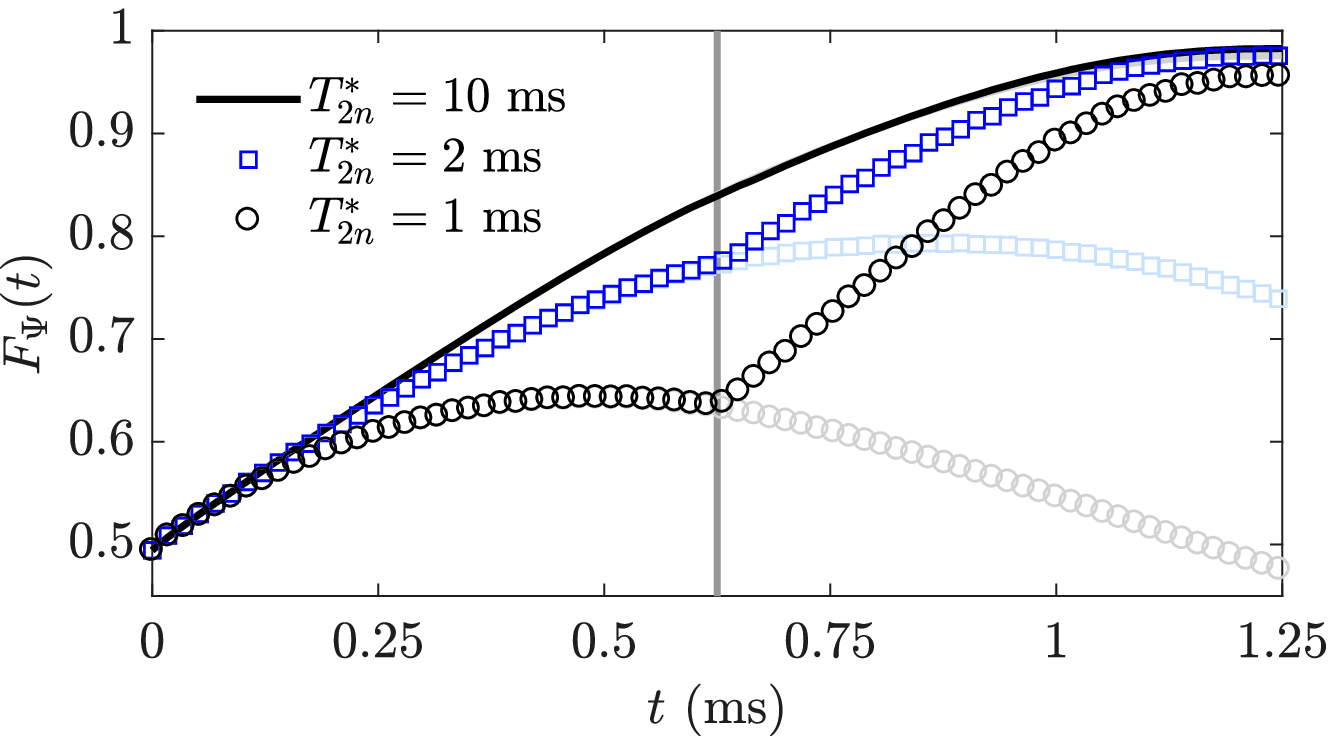}}}\hspace{2mm}
	(c)\hspace{-2mm}\vtop{\vskip+1ex\hbox{\includegraphics[width=0.95\linewidth]{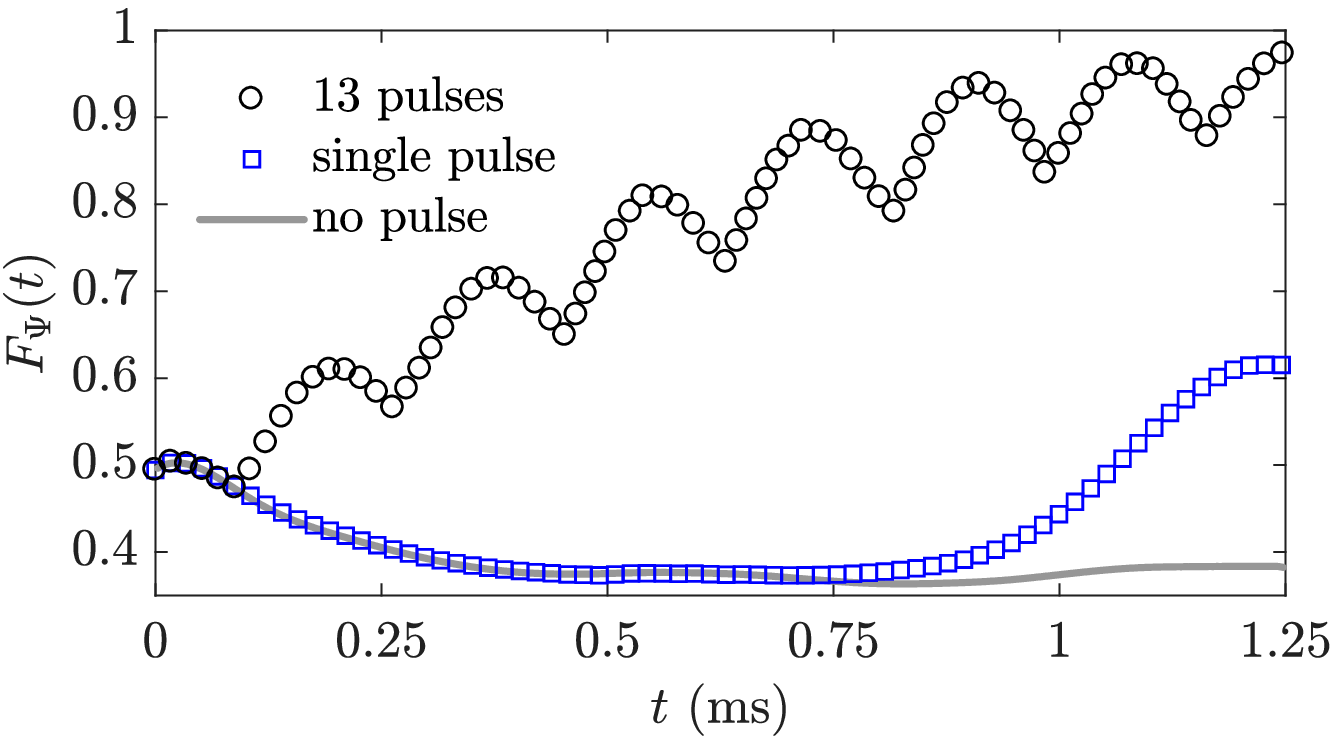}}}
	\caption{Time evolution of the fidelity $F_\Psi(t)$ under decoherence of the individual degrees of freedom. (a) Damping of the mechanical oscillator for the quality factors $Q=\{10^7,10^6,10^5\}$. (b) Dephasing of the nuclear spin for the values $T^\ast_{2n}=\{1,2,10\}\ $ms, with a single spin-echo pulse applied to both nuclear spins at $t=\pi/8J_n=0.625\ $ms (indicated by the gray vertical line). The respective gray and light blue symbols indicate the corresponding fidelity without the nuclear spin-echo pulse. (c) Dephasing of the electronic spins for the decoherence time $T^\ast_{2e}=20\ \mu$s. For comparison, the gray solid line shows the case of no nuclear echo pulse, while the blue squares and black circles show the fidelity for a single and 13 equidistant pulses, respectively. The remaining parameters are taken from Table.~\ref{tab:parameters}.}
	\label{fig:3}
\end{figure}

 Damping of the mechanical oscillator with the rate $\gamma$ by a bath at temperature $T$ can be incorporated in the system dynamics by writing the master equation~\cite{Carmichael} for the time evolution of the density operator $\rho$ according to
\begin{align}
\label{eq:mechanical_damping}
\frac{\partial}{\partial t}\rho=-i[H,\rho]+&\frac{\gamma}{2}(\bar{n}+1)\mathcal{D}[a]\rho+\frac{\gamma}{2}\bar{n}\mathcal{D}[a^\dagger]\rho,
\end{align}
with the dissipator $\mathcal{D}[O]\rho=2O\rho O^\dagger-O^\dagger O\rho-\rho O^\dagger O$ and the mean thermal occupation $\bar{n}=[\exp(-\hbar\nu/k_{\rm B}T)-1]^{-1}$. For quality factors $Q=\nu/\gamma$ of the order $Q=10^6$ and a mean occupation number of $\bar{n}=10$ we find that the mechanical dissipation has little to no effect on the nuclear spin-spin interaction, a fact that was also shown for other solid-state hybrid systems involving mechanical elements~\cite{Schuetz17,Cao17}. In current experiments with graphene oscillators and other possible candidates for nanomechanical oscillator quality factors which are even higher by some orders of magnitudes, reaching up to $Q=10^8$, can be realized~\cite{Weber14,Reinhardt16,Will17,Tsaturyan17}. As we will see, the mechanical dissipation is thereby by far the smallest source of decoherence in the effective nuclear spin-spin interaction. In fact, in the polaron picture the above expression for the mechanical damping implies an effective electron-spin pure-dephasing rate induced by the mechanical oscillators given by $\Gamma=\beta^2\gamma$~\cite{Hu15}. For $\beta=1/20$, which is the case for the parameters we consider, this leads to $\Gamma=10^{-2}\nu/4Q$, which can be exceedingly small for high-$Q$ oscillators. In order to support the claim that the mechanical damping has a very low impact on the entangling fidelity for high enough values of the quality factor, in Fig.~\ref{fig:3}(a) we show the time evolution of $F_\Psi(t)$ for the three values $Q=\{10^5,10^6,10^7\}$ with a mean thermal occupation $\bar{n}=10$. The evolution is obtained by numerical propagation of the same initial state as in Fig.~\ref{fig:2} under the master equation~\eqref{eq:mechanical_damping} with the exact Hamiltonian~\eqref{eq:H_1}. For the two values $Q=10^7$ and $Q=10^6$ there is virtually no difference in the preparation fidelity, which reaches values of $\sim99.3\%$. For lower values of the quality factor, such as $Q=10^5$, where the thermalization time $1/\gamma\approx8\ $ms of the oscillator is already on the same order as the entangling time $\pi/4J_n=1.25\ $ms, the fidelity starts to degrade. This shows that in the parameter regime we consider the mechanical damping is indeed of little importance to the nuclear spin-spin interaction for quality factors $Q\geq10^6$.

\subsection{Spin dephasing}
\label{sec:decoherence_spins}
Since we saw that the influence of the mechanical damping is negligible for high-$Q$ oscillators, the main source of decoherence is the environmental noise of the spin bath surrounding the electronic and nuclear spins of the NV centers~\cite{Dobrovitski09,Maze12}. Therefore, in order to bring out the effects of the electronic and nuclear decoherence most clearly we set $\gamma = 0$ for the time being. We consider the collective effect of the spin bath as a random shift in the energy levels of the spins since flip-flop processes can be ignored due to wide separation of the energy scales. The overall influence can be modeled by a fluctuating local magnetic field through the noise Hamiltonian
\cite{Dobrovitski09,DeLange10}
\begin{align}
H_{\rm noise}= \frac{1}{2}\sum_{i=1}^{N} \left[B_n^{(i)}(t)\tau_z^{(i)}+B_e^{(i)}(t)\sigma_z^{(i)}\right].
\end{align} 
Here, the strengths of the noise on the $i$th electronic and nuclear spins, i.e., $B_n^{(i)}(t)$ and $B_e^{(i)}(t)$, are random variables obeying a zero-mean Gaussian distribution with the autocorrelation $\langle B_\lambda^{(i)}(t) B_\lambda^{(i)}(0)\rangle = \mathcal{B}_\lambda^2\exp(-t/\tau)$ for $\lambda=e,n$. The quantities $\mathcal{B}_\lambda^2$ denote the variance of the random variables, which we assume to be equal for all $i$, and $\tau$ is the spin-bath relaxation time, which can be extracted from a measurement of the Lorentzian noise spectrum. In the following we will always assume a bath correlation time $\tau=20\ $ms. The random processes $B_\lambda^{(i)}(t)$ can be treated as Ornstein-Uhlenbeck processes~\cite{Anderson53,Klauder62,VanKampen}, and for a time discretization $\Delta t$ one can employ the following update formula of the noise strength~\cite{Gillespie,Gillespie96}:
\begin{equation}
\label{eq:update}
B_\lambda^{(i)}(t+\Delta t) = B_\lambda^{(i)}(t)e^{-\Delta t/\tau} +\mathcal{B}_\lambda\sqrt{1-e^{-2\Delta t/\tau}} n,
\end{equation}
with a normally distributed random variable $n$, which shows no temporal correlation. In all simulations presented below the evolution under the random noise was averaged over 3000 realizations of the process.

In a realistic scenario the concrete values of the noise variances $\mathcal{B}^2_\lambda$ have to be determined through free induction decay. For the exponentially decaying noise correlation and the long spin-bath correlation time that we assume, this decay has the temporal behavior $\exp(-\mathcal{B}_\lambda^2 t^2/2)$ (see Appendix~\ref{app:C}), meaning that the experimentally determined $T^\ast_{2\lambda}$ times and the noise variances stand in the relation $\mathcal{B}_\lambda=\sqrt{2}/T_{2\lambda}^\ast$. 

We first analyze the influence of the nuclear dephasing separately. To this end, we assume values of the nuclear coherence time lying in the range $T^\ast_{2n}=1-10\ $ms. Figure~\ref{fig:3}(b) depicts the time evolution of the maximally entangled state preparation fidelity for the exemplary values $T^\ast_{2n}=\{1, 2, 10\}\ $ms, corresponding to the noise strength values $\mathcal{B}_n=(2\pi)\,\{225,112.5,22.5\}\ $Hz. Without additional measures taken to counteract the dephasing of the nuclear spins we find that it is impossible to prepare the desired state, as seen by the gray and light blue makers in the second half of the plot. Since the coherence times $T_{2n}^\ast=1-2\ $ms are already on the order of the gate time $\pi/4J_n$ the noise prevents a faithful generation of an entangled state and quickly degrades the fidelity. Only for coherence times much longer than the entangling time of 1.25 ms, such as $T_{2n}^\ast=10\ $ms, can the maximally entangled state be prepared. However, even a single spin-echo pulse~\cite{Levitt} applied to the two nuclear spins at half the entangling time, i.e., at $t=\pi/8J_n=0.625\ $ms, is sufficient to overcome the influence of the dephasing noise by refocusing the nuclear state at $t=\pi/4J_n$ and thereby achieving a high fidelity, shown by the black markers in the second half of the plot. 

In the next step we investigate the influence of the electronic spin dephasing. In this case, we consider a typical value of the NV center's electronic coherence time, such as $T^\ast_{2e}=20\ \mu$s, corresponding to a noise strength of $\mathcal{B}_e=(2\pi)\,11.25\ $kHz. In Fig.~\ref{fig:3}(c) we see that a single spin-echo pulse on the nuclear spins at $t=\pi/8J_n=0.625\ $ms, as in Fig.~\ref{fig:3}(b), is not sufficient to protect the entangling operation from the noise acting on the electronic spins. Here, only a series of multiple nuclear spin-echo pulses, equidistant in time, leads to a fidelity approaching unity. Even a relatively small number of pulses, such as 13 pulses, as indicated by circles, results in a fidelity $F_\Psi(\pi/4J_n)\approx98\%$. We find that including both nuclear and electronic spin decoherences at the same time, with $T^\ast_{2n}=1\ $ms and $T^\ast_{2e}=20\ \mu$s, and applying a series of multiple equidistant echo pulses allows us to easily reach fidelities $>99\%$ for a sufficient number of pulses. We finally mention that considering merely the electron-oscillator subsystem a cooperativity can be written as $g^2 T^\ast_{2e}/\gamma(\bar{n}+1)$ and a corresponding quantity may be defined according to $A^2 T^\ast_{2e}T^\ast_{2n}$ for the electron-nuclear subsystem. Since both these values, characterizing the coherence properties of the two forms of interaction that lead to the effective nuclear spin-spin coupling, involve the electronic coherence time, it is clear that the electron dephasing has a large impact on the overall fidelity.

\section{Generation of multipartite entangled nuclear-spin states}
\label{sec:application}
In this section we apply the model for a scalable platform we introduced above to the generation of multipartite entangled states of the nuclear spins of the NV center array. Here, the Ising type of the effective interaction between the nuclear spins suggests the preparation of graph states~\cite{Briegel01}. The creation of this kind of highly entangled state as a resource for quantum computing, e.g., for the realization of a one-way quantum computer~\cite{Raussendorf03}, is in great demand. 

\subsection{Short introduction to graph states}
The aforementioned graph states correspond to mathematical graphs $G$, where in our case the vertices $V$ of the graphs are formed by the nuclear spins and the connecting edges $E$ are given by their mutual interactions. Before we show the fidelity of the graph-state preparation~\cite{Clark05,Bodiya06} using the effective nuclear spin-spin interaction we briefly review the very basics of graph states, merely for the sake of completeness; for more details we refer to Refs.~\cite{Briegel01,Hein04,Hein06}.

\begin{figure}[!b]
	\includegraphics[width=0.60\linewidth]{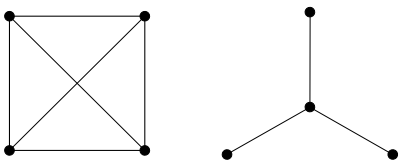}
	\caption{Example of graphs with four vertices, a complete graph (left) and a star graph (right). The circles represent the vertices and the lines the edges.}
	\label{fig:4}
\end{figure}

For a graph $G=(V,E)$ with $N$ vertices a graph state $|G_N\rangle$ is given by
\begin{align}
\label{eq:graph_states}
|G_N\rangle=\prod_{\{i,j\}\in E}\mathcal{U}_{ij}|+\rangle^{\otimes N},
\end{align}
where $i$ and $j$ label the vertices of the graph, which are connected by an edge $\{i,j\}$, and $\mathcal{U}_{ij}$ denotes a phase gate between them. Furthermore, we used the product-state notation $|\psi\rangle^{\otimes N}=|\psi\rangle_1|\psi\rangle_2\cdots|\psi\rangle_N$ for a graph $V$ consisting of $N$ vertices whose states are labeled by the subscript. For example, it was shown~\cite{Hein04,Hein06} that the graph states corresponding to a complete graph and a star graph (see Fig.~\ref{fig:4}) are equivalent to Greenberger-Horne-Zeilinger (GHZ) states~\cite{Greenberger89}, which show a maximal violation of the Bell inequalities and are defined as $|{\rm GHZ}_N\rangle=[\up^{\otimes N}+\down^{\otimes N}]/\sqrt{2}$. 

Up to local terms that commute with the interaction, the effective Hamiltonian~\eqref{eq:H_eff} is equivalent to the form
\begin{align}
\label{eq:graph_Hamiltonian}
H_{\rm graph}=4J_n\sum_{i<j}^N\frac{1+\tau_z^{(i)}}{2}\frac{1-\tau_z^{(j)}}{2},
\end{align}
which is used in the seminal paper~\cite{Briegel01} on graph states. The time evolution under this Hamiltonian for the evolution time $t_g=\pi/4J_n$ exactly generates the unitaries $\mathcal{U}_{ij}$ from the definition of the graph states given in Eq.~\eqref{eq:graph_states}. This means that if initially the nuclear spins are prepared in the state $|+\rangle^{\otimes N}_n$, after the time $t_g$ the state has evolved into a graph state. Since in our model all nuclear spins effectively interact with each other, the generated graph state is naturally a complete graph state and thereby equivalent to the maximally entangled GHZ state. 

\subsection{Preparation of nuclear-spin graph states with three and four vertices}
Since our focus lies on the nuclear degrees of freedom, in the remainder of this section we abbreviate the state of the $i$th nuclear spin according to $|\psi\rangle_n^{(i)}=|\psi\rangle_i$. For an array of three NV centers, corresponding to a complete graph with three vertices, after the time $t_g$ the ideal Hamiltonian~\eqref{eq:graph_Hamiltonian} would transform the initial nuclear state $|+\rangle^{\otimes 3}$ into the graph state
\begin{align}
|G_3\rangle=\frac{1}{2}\Big[&\Big(\up_1\up_2-\down_1\down_2\Big)|-\rangle_3\nonumber\\
-&\Big(\up_1\down_2+\down_1\up_2\Big)|+\rangle_3\Big],
\end{align}
while for $N=4$ the initial states $|+\rangle^{\otimes 4}$ become the graph state
\begin{align}
    |G_4\rangle=\frac{1}{\sqrt{8}}\Big[&\Big(\up_1\up_2-\down_1\down_2\Big)\nonumber\\
    &\times\Big(\up_3|+\rangle_4 +\down_3|-\rangle_4\Big)\nonumber\\
    -&\Big(\up_1\down_2+\down_1\up_2\Big)\nonumber\\
    &\times\Big(\up_3|-\rangle_4 +\down_3|+\rangle_4\Big)\Big].
\end{align}
As mentioned in the previous section, these two states are unitarily equivalent to the states $|{\rm GHZ}_3\rangle$ and $|{\rm GHZ}_4\rangle$ and thereby maximally entangled. 

In an ideal scenario, without any kind of noise, the exact Hamiltonian~\ref{eq:H_1} allows us to prepare these graph states with a very high fidelity. Not surprisingly, the situation changes drastically when the random noise on the spins is taken into account, where for realistic dephasing times the preparation fidelity merely shows values of around $50\%$. As pointed out earlier, a continuous dynamical decoupling with a radio-frequency driving field to protect the nuclear-spin coherence would imply an additional term $\sum_{i=1}^N\Omega_n^{(i)}\tau_x^{(i)}/2$ in the effective nuclear Hamiltonian~\eqref{eq:H_eff}, which clearly does not commute with the effective Ising interaction and thereby leads to a nonequivalence to the graph-state Hamiltonian~\eqref{eq:graph_Hamiltonian}. Therefore, in order to still make the creation possible, we employ a pulsed decoupling, which we showed in Sec.~\ref{sec:decoherence_spins} to work well for the entanglement of two NV center nuclear spins. The time evolution of the initial state $|+\rangle^{\otimes N}$ under the graph-state Hamiltonian, but including a series of instantaneous $\pi$ pulses results in a state $|\tilde{G}_N\rangle$ that is equivalent to the graph states $|G_N\rangle$ up to a local unitary transform, thus showing the same degree of entanglement.

\begin{figure}[!t]
	\centering
	(a)\hspace{-2mm}\vtop{\vskip+1ex\hbox{\includegraphics[width=0.95\linewidth]{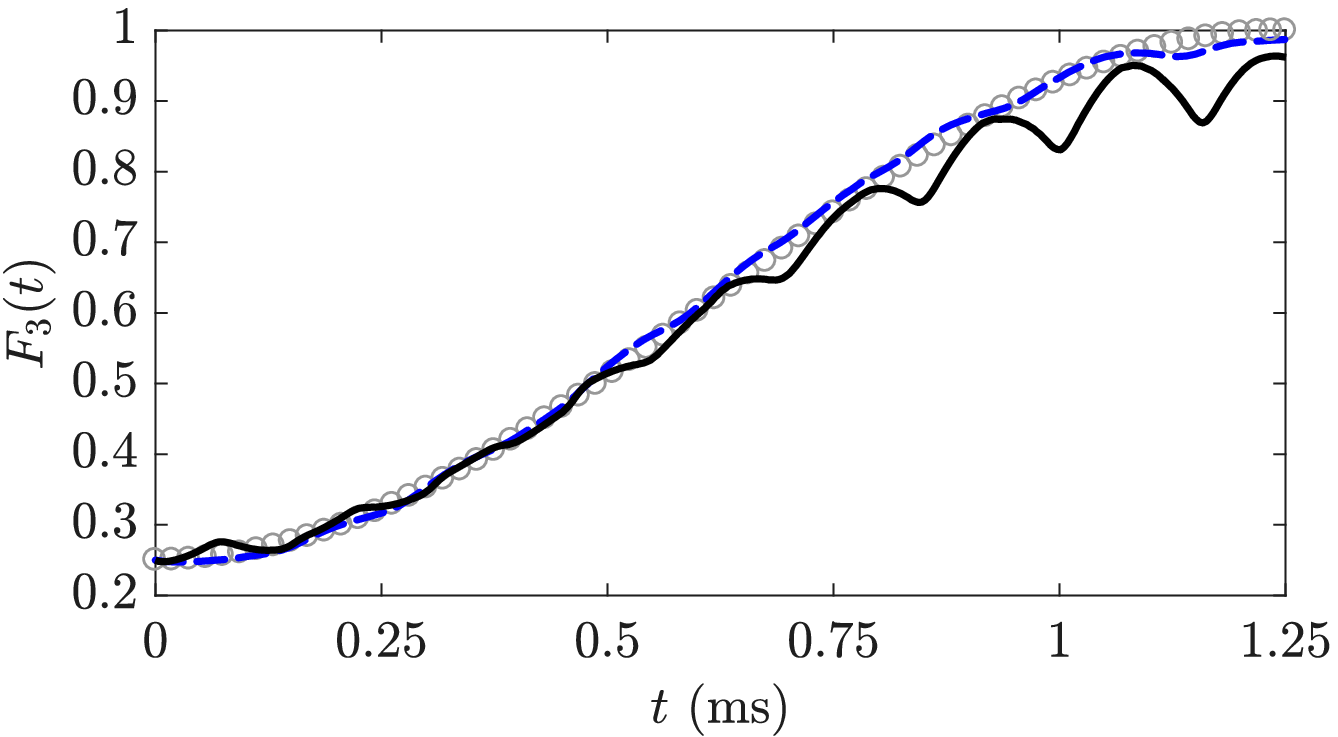}}}\hspace{2mm}
	(b)\hspace{-2mm}\vtop{\vskip+1ex\hbox{\includegraphics[width=0.95\linewidth]{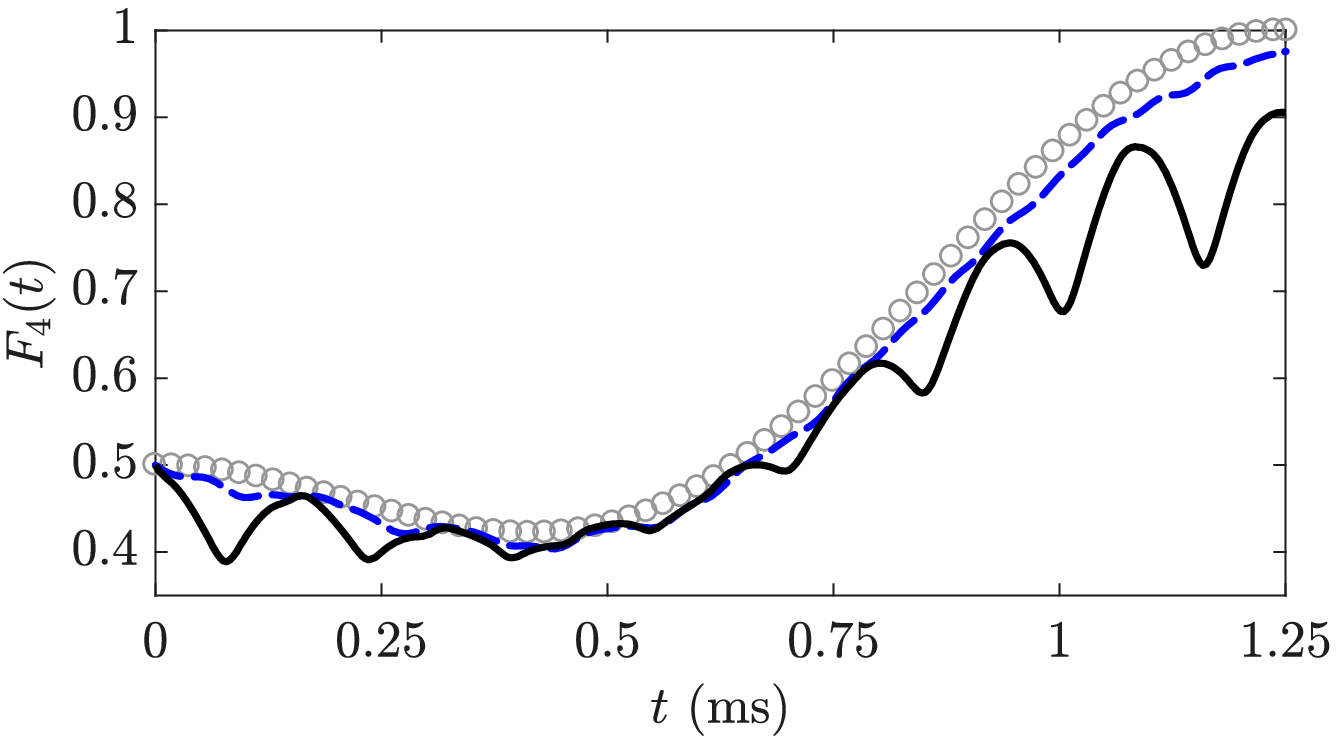}}}
	\caption{Fidelity $F_N(t)$ of the graph-state preparation using 15 temporally equidistant decoupling pulses on all nuclear spins for: (a) $N=3$ and (b) $N=4$. For comparison, the gray circles show the fidelity of a perfect preparation using the graph-state Hamiltonian~\eqref{eq:graph_Hamiltonian}. The blue dashed lines represent a preparation using the exact Hamiltonian~\eqref{eq:H_1}, and the black solid lines show the corresponding fidelity including spin dephasing with the coherence times $T_{2n}^\ast=1\ $ms and $T^\ast_{2e}=20\ \mu$s. The remaining parameters are taken from Table~\ref{tab:parameters}.}
	\label{fig:5}
\end{figure}

In order to analyze the fidelity of the graph-state preparation for $N$ vertices using decoupling pulses we define
\begin{align}
F_N(t)=\langle \tilde{G}_N|{\rm Tr}_{e-{\rm osc}}\{\rho(t)\}|\tilde{G}_N\rangle,
\end{align}
where the density operator $\rho(t)$ is obtained by numerical propagation, including the instantaneous $\pi$ pulses, under the exact Hamiltonian~\eqref{eq:H_1} with the nuclear initial states $|+\rangle^{\otimes N}$, while all electronic spins are prepared in the state $|+\rangle_e$ and the oscillator is in its ground state. Figure~\ref{fig:5} shows this fidelity $F_N(t)$ for the two above-mentioned cases, i.e., $N=3$ and $N=4$, where 15 instantaneous $\pi$ pulses, equidistant in time, were applied to all nuclear spins. 
As a reference, in both Figs.~\ref{fig:5}(a) and~\ref{fig:5}(b) the gray circles indicate the evolution of the fidelity without noise for a preparation using the graph-state Hamiltonian~\eqref{eq:graph_Hamiltonian} and decoupling pulses. Here, by definition, the target graph state $|\tilde{G}_N\rangle$ is prepared with a unit fidelity. The blue dashed lines, on the other hand, represent the fidelity $F_N(t)$ when the exact Hamiltonian~\eqref{eq:H_1} is used, still in an ideal scenario without noise. Finally, in both cases we included a noise on the spins resulting in the coherence times $T_{2n}^\ast=1\ $ms and $T^\ast_{2e}=20\ \mu$s. The pulses prove to be effective also for the case of multiple centers and are able to achieve good preparation fidelities in the presence of strong spin dephasing. After a successful generation of the entangled state the effective nuclear interaction can then be switched off by a sudden ramp up of the electronic Rabi frequency in order to diminish $J_n$.

\section{Conclusion}
\label{sec:conclusion}
We have demonstrated that the combination of an engineered array of NV centers, e.g., created by ion-implantation techniques, with a suitably tailored mechanical oscillator element, can build a hybrid platform in which an effective all-to-all Ising interaction between the nuclear spins of the NV centers is present. This interaction is mediated by the electronic spins, whose long-range interaction is established by the mechanical oscillator. The spatial range of the oscillator-mediated long-range interaction is thereby given by the mode function of the mechanical oscillation, providing the possibility of a scalable array without strict conditions on the inter-NV center distance. 

The analysis of the decoherence of the different components showed that the mechanical damping has by far the smallest impact on the effective nuclear-nuclear interaction and for high-$Q$ oscillators, such as $Q\geq10^6$, may safely be neglected in the time intervals of interest. The spin dephasing, modeled by a random magnetic field, on the other hand, leads to a tremendous degradation of the applicability if no countermeasures using decoupling techniques are taken. The continuous microwave driving field together with a sequence of Hahn-echo pulses on the nuclear spins with relatively small numbers of $\pi$ pulses leads to high fidelities for the creation of maximally entangled states of multiple nuclear spins.

\begin{acknowledgments}
This work is supported by the National Key R\&D Program of China (Grant No. 2018YFA0306600), the National Natural Science Foundation of China (Grants No. 11574103, No. 11690030, No. 11690032), the China Postdoctoral Science Foundation (Grant No. 2017M622398), and the National Young 1000 Talents Plan. The authors thank Y. Chu, Y. Liu, and Prof. M. B. Plenio for helpful discussions.
\end{acknowledgments}

\appendix
\section{Polaron transform of the electronic driving field}
\label{app:A}
The exact polaron transformation of the driving field term of the electronic spins reads
\begin{align}
\mathcal{P}\Bigg[\sum_{i=1}^N\frac{\Omega_i}{2}\sigma_x^{(i)}\Bigg]\mathcal{P}^\dagger= \sum_{i=1}^N\frac{\Omega_i}{2}D(2\alpha_i)\sigma_+^{(i)}+{\rm H.c.},
\end{align}  
with the usual electronic raising and lowering operators $\sigma_\pm^{(i)}=(\sigma_x^{(i)}\pm i\sigma_y^{(i)})/2$. In the interaction picture with respect to the remaining terms of the Hamiltonian $H_\mathcal{P}$, the terms of the displacement operator involving different powers of annihilation and creation operators, such as $a^na^{\dagger m}$, with positive integers $n$ and $m$, rotate with the frequency $(m-n)\nu$. Their magnitude, on the other hand, is on the order of $\alpha_i^{n+m}\Omega_i$. This implies that in the parameter regime we consider, a rotating-wave approximation can be performed, and all rotating terms can be safely discarded, which was also verified by numerical comparison. The surviving terms are the contributions containing equal powers of $a$ and $a^\dagger$, which are given by the diagonal elements of the displacement operator in the number-state basis~\cite{Cahill69}, in our case yielding the approximation
\begin{align}
D(2\alpha_i) \approx e^{-2\alpha_i^2}L_{a^\dagger a}(4\alpha_i^2),
\end{align}
with the Laguerre polynomials $L_n(x)$. For a ground-state cooled mechanical oscillator we can then set $a^\dagger a=0$ and use $L_0(x)=1$, leading to 
\begin{align}
\mathcal{P}\Bigg[\sum_{i=1}^N\frac{\Omega_i}{2}\sigma_x^{(i)}\Bigg]\mathcal{P}^\dagger\approx \sum_{i=1}^N \frac{\Omega_ie^{-2\alpha_i^2}}{2}\sigma_x^{(i)},
\end{align}
which was used in the main text.\vspace{-2ex}

\section{Schrieffer-Wolff transform of the nuclear driving field}
\label{app:B}
The transformation of a nuclear driving field 
\begin{align}
H_{nd}=\sum_{i=1}^N\frac{\Omega_n^{(i)}}{2}\tau_x^{(i)}
\end{align}
is given by
\begin{align}
\mathcal{S}\Bigg[\sum_{i=1}^N\frac{\Omega_n^{(i)}}{2}\tau_x^{(i)}\Bigg]\mathcal{S}^\dagger=&\sum_{i=1}^N\Bigg[(1-2\beta_i^2)\frac{\Omega_n^{(i)}}{2}\tau_x^{(i)}\nonumber\\
&+\beta_i\Omega_n^{(i)}\sigma_y^{(i)}\tau_y^{(i)}+\mathcal{O}(\beta_i^3)\Bigg],
\end{align}
where we already see that the nuclear Rabi frequency is also renormalized according to $\bar\Omega_n^{(i)}=(1-2\beta_i^2)\Omega_n^{(i)}$. On the other hand, the $\sigma_y^{(i)}\tau_y^{(i)}$ term in the second line could be safely discarded by making another rotating-wave approximation. \vspace{5ex}

\section{Simulations of the random noise}
\label{app:C}
For the update formula~\eqref{eq:update} of the random noise process we check whether the numerical implementation fulfills the required properties. For all following simulations we averaged over 3000 realizations, which proved to be a sufficiently large sample number.
\begin{figure}[b]
	\centering\vspace{0ex}
	(a)\hspace{-2mm}\vtop{\vskip+1ex\hbox{\includegraphics[width=0.92\linewidth]{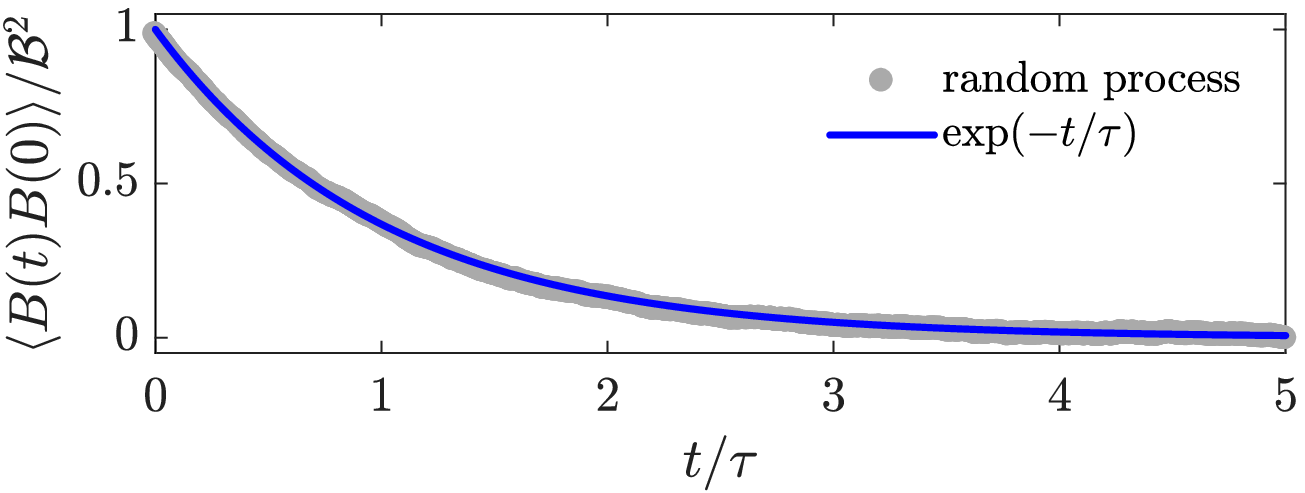}}}\hspace{2mm}
	(b)\hspace{-2mm}\vtop{\vskip+1ex\hbox{\includegraphics[width=0.92\linewidth]{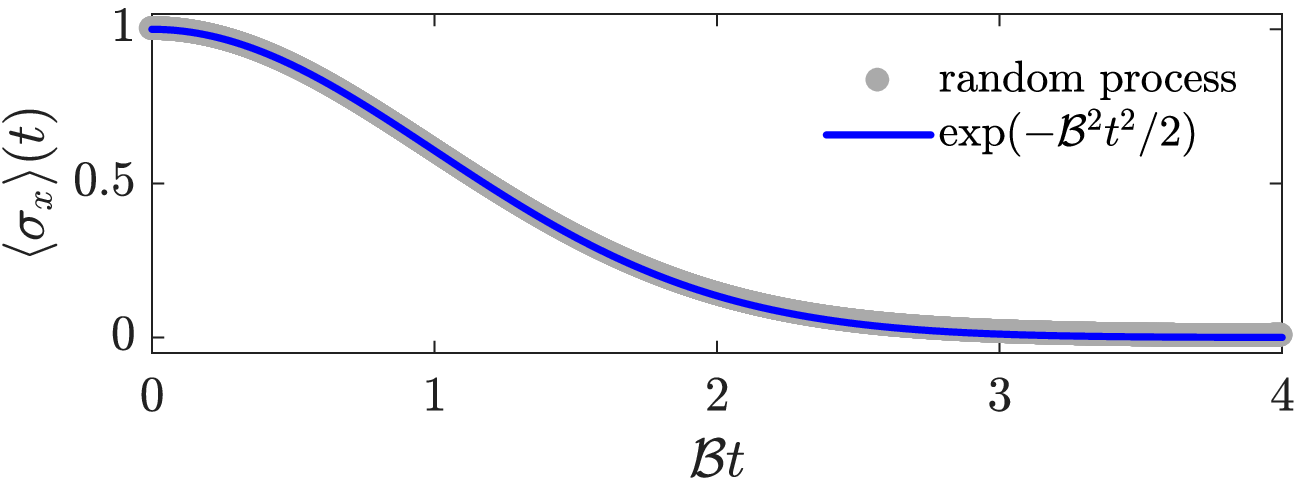}}}
	\caption{(a) Simulation of the noise autocorrelation function verifying its exponential decay. (b) Free induction decay simulation with $\mathcal{B}\tau=\sqrt{2}\times1000$, verifying the Gaussian decay. In both simulations the average is taken over 3000 realizations of the random process.}
	\label{fig:6}
\end{figure}
In Fig.~\ref{fig:6}(a) we show the noise autocorrelation function, which nicely reproduces the exponential decay. Figure~\ref{fig:6}(b) shows a simulation of the free induction decay of a single qubit, i.e., the time evolution of an initial superposition state $|+\rangle$ under the noise Hamiltonian $B(t)\sigma_z/2$. As a relation between the noise variance and the correlation time we took $\mathcal{B}\tau=\sqrt{2}\times1000$. The decay shows the well-known Gaussian decay, which in experiments is used to determine the $T_2^\ast$ time according to $\langle\sigma_x\rangle(t)=\exp(-t^2/T^{\ast 2}_2)$. This allows us to identify $\mathcal{B}=\sqrt{2}/T^\ast_2$, which for typical parameters of NV center electron spins, viz., $\tau=20\ $ms and $T^\ast_{2e}=20\ \mu$s, yields a strength of the electronic noise $\mathcal{B}_e\approx(2\pi)\, 11.25\ $kHz, while for nuclear coherence times $T^\ast_{2n}=1\ $ms this leads to $\mathcal{B}_n=\mathcal{B}_e/50\approx(2\pi)\,0.225\ $kHz.


%

\end{document}